\documentclass[amsmath,amssymb,aps,prl,twocolumn]{revtex4-2}
\usepackage{graphicx}
\usepackage{bm}
\usepackage{array}
\usepackage{amsmath}
\usepackage{contour}
\usepackage{physics}
\usepackage[english]{babel}
\usepackage{float}
\usepackage{nicematrix}
\usepackage[final]{hyperref}
\hypersetup{
	colorlinks=true,       
	linkcolor=blue,        
	citecolor=blue,        
	filecolor=magenta,     
	urlcolor=blue         
}
\usepackage{tikz}

\begin{document}
	\title{Signatures of exceptional points in multiterminal \\ superconductor-normal metal junctions}
	
	\author{Oliver Solow}
	\affiliation{Center for Quantum Devices, Niels Bohr Institute, University of Copenhagen, DK-2100 Copenhagen, Denmark}
	\author{Karsten Flensberg}
	\affiliation{Center for Quantum Devices, Niels Bohr Institute, University of Copenhagen, DK-2100 Copenhagen, Denmark}
	\date{\today}
	\begin{abstract}
		We study a non-Hermitian, multiterminal superconducting-normal system in order to identify experimental signatures of exceptional points. We focus on a minimal setting with a single spinful level, spin-dependent normal leads, and a noncollinear magnetic field. This system hosts both topologically-protected, as well as symmetry-protected exceptional points. Using an exact transport formalism, we show that the exceptional points leave signatures visible through spectroscopy of the Andreev states, but that they have a minor effect on the Josephson current. We also argue that these findings hold with interactions.
	\end{abstract}
	\maketitle
	
	A distinct difference between closed and open quantum systems is the possibility of open systems to have exceptional points (EPs) where the non-Hermitian Liouvillian matrix is non-diagonalizable. The consequences of such points in parameter space is an active research area in, for example, photonics and superconducting qubits \cite{san-jose_majorana_2016,ashida_non-hermitian_2020, wang_non-hermitian_2023, zhou_observation_2018,ochkan_non-hermitian_2024,bergholtz_exceptional_2021}. In the context of condensed matter systems, the experimental consequences have mostly been discussed in terms of post-selection \cite{minganti_hybrid-liouvillian_2020,minganti_quantum_2019,molmer_monte_1993} which, however, is experimentally costly due to the large number of repetitions required \cite{naghiloo_quantum_2019}. More recently, exceptional points and their experimental consequences in multiterminal Josephson junctions have attracted theoretical interest\cite{cayao_non-hermitian_2024,cayao_non-hermitian_2024-1,li_anomalous_2024,ohnmacht_non-hermitian_2024,shen_non-hermitian_2024, capecelatro_andreev_2025,pino_thermodynamics_2025}, including the topological properties of full counting statistics\cite{javed_fractional_2023,pavlov_topological_2025}. Another recent and interesting direction is the generalization of the famous ten-fold topological classification \cite{altland_nonstandard_1997} of non-interacting Hamiltonians to the non-Hermitian case. Here, it turns out that instead of ten classes, thirty-eight classes are needed \cite{kawabata_symmetry_2019}. 
	
\begin{figure*}[t]
		\centering
		\includegraphics[width=\textwidth]{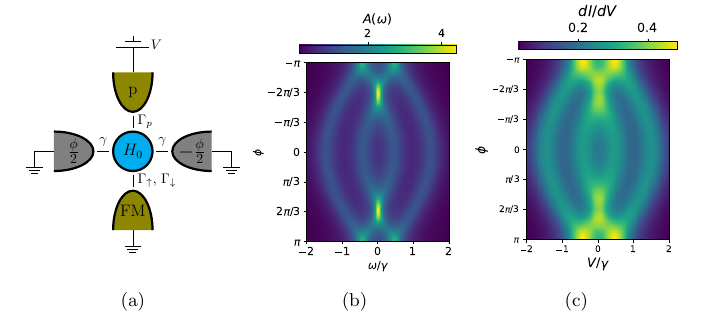}
		\caption{(a) Schematic of the multi-terminal SNS setup. The gray leads are superconductors, the FM lead is ferromagnetic, and the p lead is the probe used for spectroscopy. (b) Spectral function of the central region of a single-level system. (c) $\frac{dI}{dV}$ calculation for the same system at $T=0.1\gamma$, demonstrating the same qualitative features as the spectral function. In particular, the EP shows up in both as branching points. The parameters for (a) and (b) are $\theta=\frac{\pi}{3}$, $\varepsilon=0$, $B=0.5\gamma$, $\Gamma_\uparrow=\gamma$ and $\Gamma_\downarrow=0$}
		\label{fig:Schematic}
		\label{fig:Conceptual}
	\end{figure*}

	In this paper, we study the experimental signatures of such exceptional points in the context of a dissipative SNS junction where the junction is a normal system also coupled to a normal lead (see Fig.~\ref{fig:Schematic}(a)). The normal region supports a single broadened Andreev level whose propagator can have exceptional points. We show that these points leave signatures in spectroscopic measurements, but also that there are no clear features showing the appearance of exceptional points in the thermodynamical properties such as supercurrent, which has been a topic of recent debate in the literature\cite{cayao_non-hermitian_2024,shen_non-hermitian_2024,beenakker_josephson_2024}.
	
	To understand the properties of these multiterminal SNS junction, we consider the Green function of the central region (see Fig.~\ref{fig:Schematic}(a)) $G(\omega)=[\omega-H_0+\Sigma(\omega)]^{-1}$, where $H_0$ is the Hamiltonian of the central region and $\Sigma(\omega)$ is the self-energy induced by the leads. For simplicity, we start by considering the infinite-gap limit, and later generalize to a finite superconducting gap. In this limit, the self-energy $\Sigma(\omega)$ is $\omega$-independent, and thus we can write $H_0-\Sigma(\omega)\approx H_0-\Sigma(0)=H_\mathrm{eff}$, where $H_\mathrm{eff}$ is often denoted as the effective Hamiltonian of the system. Since $\Sigma(\omega)$ is generally non-Hermitian, so is $H_\mathrm{eff}$. While it is natural to treat $H_\mathrm{eff}$ as a Hamiltonian, the fact that it is non-Hermitian leads to ambiguities when defining observables. To avoid such ambiguities we use exact transport formalism (valid for non-interacting leads) to derive predictions for the experimental observable. Later we discuss how these findings generalize to the interacting case.
	
	The system shown in Fig.~\ref{fig:Schematic}(a) is a four-terminal SNS junctions with a central region coupled to two superconducting leads and a spin-polarized normal lead, giving rise to the dissipative behavior. In addition, we add a weak coupling to a probe lead which is used to probe the structure of the normal region and its behavior due to the exceptional points. This setup admits several interesting scenarios, depending on the details of the central region and the coupling to the dissipative lead. We focus on the simplest possible case (which is easily generalized) where the central region has a single level with energy $\varepsilon$, with a spin-dependent coupling $\Gamma_{\uparrow/\downarrow}$ to the normal lead, and a magnetic field aligned at an angle $\theta$ non-collinear with the polarization of the normal lead. Starting our discussion in the infinite-gap limit, we can write down the effective Hamiltonian
	\begin{equation}
		H_\mathrm{eff}=\begin{pmatrix}
			\varepsilon_\uparrow & B_x & \gamma\cos{\frac{\phi}{2}} & 0\\
			B_x & \varepsilon_\downarrow & 0 & \gamma\cos{\frac{\phi}{2}}\\
			\gamma\cos{\frac{\phi}{2}} & 0 & -\varepsilon_\downarrow^* & B_x\\
			0 & \gamma\cos{\frac{\phi}{2}} & B_x & -\varepsilon_\uparrow^*
		\end{pmatrix}
		\label{hamiltonian}
	\end{equation}
	where $\varepsilon_{\uparrow/\downarrow}=\varepsilon\pm B\cos{\theta} + i\Gamma_{\uparrow/\downarrow}$ and $B_x=B\sin{\theta}$. The case of $\theta=0$ has been studied before in \cite{cayao_non-hermitian_2024}, which found exceptional points present for $\varepsilon=0$. Here we point out that the general case $\theta\neq0$ allows for different classes of exceptional point in the same system. 
	
\begin{figure*}[t!]
		\includegraphics[width=\textwidth]{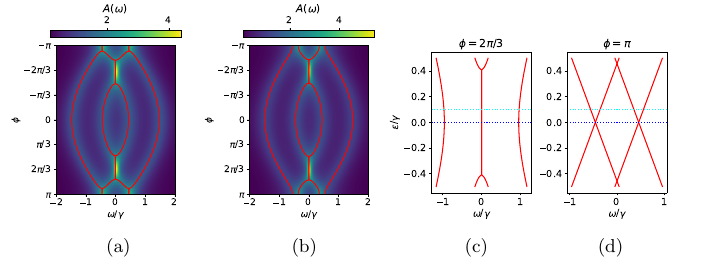}
		\caption{Spectral function computations for single-level system. The red lines marks the real part of the poles of the spectral function.(a) Spectral function for $\theta=\frac{\pi}{3}$ and $\varepsilon=0$, showing both the fragile and robust exceptional points (b) Spectral function for the same $\theta$ with $\varepsilon=0.1$, demonstrating the disappearance of the fragile exceptional points (c-d) Energy as a function of $\varepsilon$ at $\phi=\frac{2\pi}{3}$(c) and $\phi=\pi$(d). The degeneracy of the real part of the poles for a large range of parameters at $\phi=2$ marks a topologically distinct phase from the non-degenerate case in (c), whereas in (d) the degeneracy is not topologically protected. The blue and cyan lines mark the values of $\varepsilon$ for (a) and (b), respectively. Except where otherwise noted, parameters are the same as in Fig.~\ref{fig:Schematic}.}
		\label{fig:Spectral-examples}
		\label{fig:both-points}
		\label{fig:only-robust}
		\label{fig:stability}
	\end{figure*}

	In the weak probe limit, the conductance $\frac{dI}{dV}$ is proportional to the spectral function $A(\omega)$ (up to thermal broadening), as is clearly seen in Fig.~\ref{fig:Schematic}. Thus, by studying the spectral function, we get a better understanding of the probe differential conductance, which is one of the main observables of our setup. Looking at the real part of the poles, the exceptional points are visible in the spectroscopy plot as the points where the real part of the poles merge or split (see figs. \ref{fig:both-points}(a-b)). 
	However, we emphasize that this merger of lines is not a conclusively signature of the presence of exceptional points, other effects could in principle give similar looking spectroscopy data.
	
	By varying the parameters of the system, we find that there are two types of exceptional points: ``fragile'' exceptional points which are only present when $\varepsilon=0$, and ``robust'' exceptional points which are present for a large range of parameters (see Fig.~\ref{fig:stability}(c)).
	
	The difference between the fragile and robust exceptional points can be explained by their different symmetry properties, as understood using the 38-fold way of classifying non-Hermitian topology\cite{kawabata_symmetry_2019}. Since each exceptional point involves two poles of $G(\omega)$ and thus two eigenvalues of $H_\mathrm{eff}$, we will study the exceptional points by projecting $H_\mathrm{eff}$ onto the two-dimensional subspace where the exceptional point is present. To find the robust exceptional points, we start by performing a rotation such that the $z$-axis is aligned with $\vec{B}$ instead of with the normal lead polarization, thus making the dissipation off-diagonal(see appendix). In this form, we can project onto the relevant subspace to obtain the Hamiltonian
	\begin{equation}
		H_{\mathrm{robust}}=\begin{pmatrix}
			\varepsilon_1(\theta,\phi) & ig(\theta,\phi)\\ ig^*(\theta,\phi) &-\varepsilon_1^*(\theta,\phi)
		\end{pmatrix}.
		\label{robust}
	\end{equation}
	The robust Hamiltonian obeys PHS$^\dagger$ symmetry, which places it in the class D$^\dagger$. Indeed, any open BdG system obeys PHS$^\dagger$ symmetry. For a zero-dimensional system with a point gap, this class has a $\mathbb{Z}_2$ topological invariant. To find the topological invariant, we rewrite the Hamiltonian as $H=Iz_{gap}+\vec{d}\cdot\vec{\sigma}$\cite{budich_symmetry-protected_2019} where $z_{gap}$ is a complex number, and $\vec{d}$ is a complex vector. We can then define the topological invariant as 
	\begin{equation}
		w=\text{sign}\left(\det(\vec{d}\cdot\vec{\sigma})\right).
	\end{equation}
	For the robust Hamiltonian, the topological invariant is given by
	\begin{equation}
		w_\mathrm{robust}=\text{sign}(\abs{g}^2-\Re(\varepsilon_1)^2).
	\end{equation}
	Here, $g$ is the dissipation strength in the subspace and basis described in appendix \ref{robust-EP}. It depends non-trivially on $\phi$, but importantly $|g|$ can take values both larger and smaller than $\varepsilon_1$ for a large range of parameters.
	
	For the fragile exceptional points, the Hamiltonian can again be projected onto the relevant subspace. This is particularly easy to see in the case of $\theta=0$, since in this case (\ref{hamiltonian}) decouples into two blocks, each of the form
	\begin{equation}
		H_{\mathrm{fragile}}=\begin{pmatrix}
			\varepsilon_1 & \gamma\cos{\frac{\phi}{2}}\\ \gamma\cos{\frac{\phi}{2}} &-\varepsilon_2^*
		\end{pmatrix}.
	\end{equation}
	The fragile Hamiltonian does not obey any relevant symmetries, and thus falls in class A which in this case has no topological invariants\cite{ohnmacht_non-hermitian_2024}. However, if we perform the same rewriting as for the robust case, we see that when $\Re(\varepsilon_1)=-\Re(\varepsilon_2)$, the matrix $\vec{d}\cdot\vec{\sigma}$ obeys TRS and PHS, and thus falls into the class BDI, giving it a $\mathbb{Z}_2$ topological invariant
	\begin{equation}
		w_\mathrm{fragile}=\mathrm{sign}\left(\Im(\varepsilon_1-\varepsilon_2)^2-\gamma^2\cos^2\left(\frac{\phi}{2}\right)\right).
	\end{equation}
	Since TRS and PHS are contingent on a particular symmetry, namely that $\Re(\varepsilon_1)=-\Re(\varepsilon_2)$, we see that this Hamiltonian displays a symmetry-protected topological phases.
	For non-Hermitian systems with a $\mathbb{Z}_2$ topological invariant, the Hamiltonian has exceptional points at the boundary between the two topological phases in parameter space. Thus, the signatures of exceptional points are related to the signatures of topological phases in the system. We will now discuss these signatures in the observable quantities.
	\begin{figure}[!]
		\centering
		\includegraphics[width=0.8\linewidth]{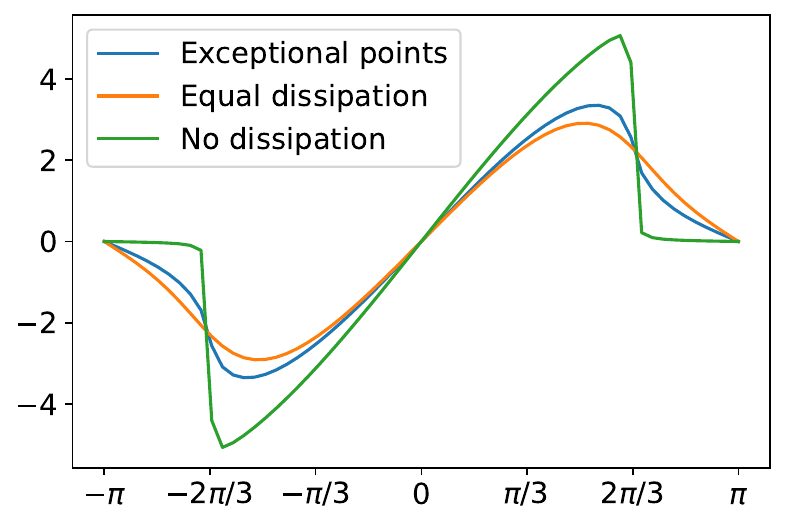}
		\caption{Josephson current for a multiterminal junction with $\Gamma_\uparrow=\gamma, \Gamma_\downarrow=0$, which has an exceptional point, $\Gamma_\uparrow=\Gamma_\downarrow=0.5\gamma$, which does not have an exceptional point, and $\Gamma_\uparrow=\Gamma_{\downarrow}=0$, which also does not have an exceptional point. We observe no distinct features in the case with exceptional points. All other parameters are as in Fig.~\ref{fig:Schematic}.}
		\label{fig:Josephson-current}
	\end{figure}
	There are two primary observables of this system: The probe conductance, which as discussed is related to the spectral function, and the Josephson current between the two superconducting leads. We will consider each in turn.
	Considering the spectral functions in Figs.~\ref{fig:Conceptual} and~\ref{fig:Spectral-examples}, we see that the poles of the spectral function yield clear lines, and we see that the exceptional points lead to the lines merging and splitting. The exceptional points indicate the phase boundary between two different phases: one with a real gap between the relevant eigenvalues, and one with an imaginary gap. In the phase with a real gap, the gap is simply the splitting between the two lines. In the phase with an imaginary gap, the imaginary parts of the corresponding roots of $G(\omega)$ split, with one root acquiring a larger imaginary component and one root acquiring a smaller imaginary component. This corresponds to one line becoming wider and the other becoming narrower, and this line narrowing is visible in the spectral function, and is thus a sign that the system is in the imaginary-gap phase.
	\begin{figure*}
		\centering
		\includegraphics[width=\textwidth]{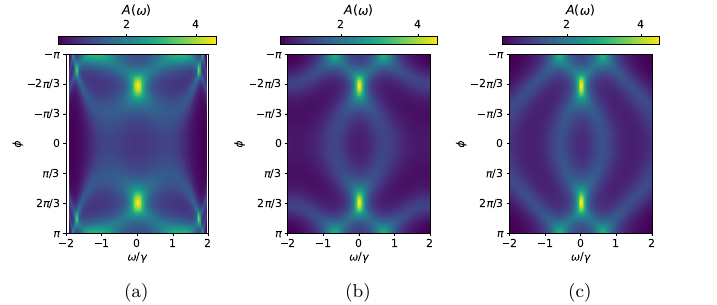}
		\caption{Spectral function of the junction with finite $\Delta$. Subfigures (a), (b) and (c) correspond to $\Delta/\gamma=2,3$ and $4$ respectively, with all other parameters as in Fig.~\ref{fig:Schematic}. We see that as $\Delta$ increases, the qualitative features quickly approach those for the $\Delta\rightarrow\infty$ case.}
		\label{fig:Finite-D}
	\end{figure*}
	There has been debate on the effect of exceptional points on the Josephson current of the system. While it has been claimed that the presence of exceptional points enhances Josephson currents\cite{cayao_non-hermitian_2024}, this claim is based on treating $H_\mathrm{eff}$ as a proper Hamiltonian, which leads to ambiguities in defining the Josephson current\cite{shen_non-hermitian_2024,pino_thermodynamics_2025}. When using the correct transport formalism, this ambiguity disappears. To study the effect of exceptional points on the Josephson current, we consider three different setups: A system with an exceptional point due to dissipation $\Gamma_{\uparrow}$ in one spin channel, a system which has dissipation $\Gamma=\frac{\Gamma_\uparrow}{2}$ in both spin channels and no exceptional point, and a system with no dissipation (see Fig.~\ref{fig:Josephson-current}). We see that the system with exceptional points show an enhancement of the Josephson current compared to the system with equal dissipation, but no enhancement compared to the system with no dissipation. This indicates that Josephson current enhancement is not a universal feature of exceptional points, and that spectroscopy is a better method for probing their presence.
	
	With the transport formalism, we can explore the spectral function and associated differential conductance outside the infinite-gap limit (see Fig.~\ref{fig:Finite-D}). We see that for a gap which is finite, but still the largest energy scale of the system, the spectral function is qualitatively similar to the infinite-gap limit, and displays the same exceptional points. This indicates that the exceptional points will be visible in realistic experimental setups.
	
\begin{figure*}
		\centering
		\includegraphics[width=\textwidth]{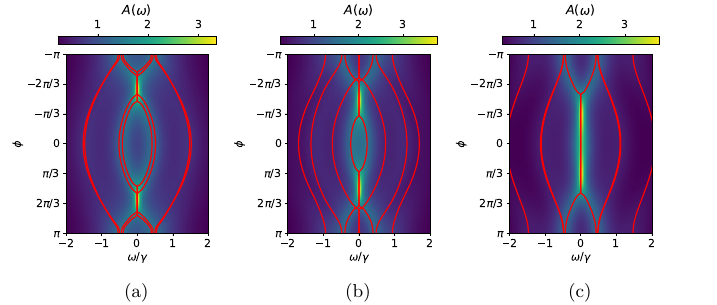}
		\caption{Spectral function for junction with finite $U$, with red lines corresponding the poles of the Green function found by the PERLind method. Subfigures (a), (b) and (c) correspond to $U/\gamma=0.1,0.5$ and $1$ respectively, with all other parameters as in Fig.~\ref{fig:Schematic}. We see that the robust exceptional points are present for both $U=0.1 \gamma$ and $U=0.5\gamma$, and that for $U=\gamma$ two of the robust points merge. }
		\label{fig:U}
	\end{figure*}

	One way to construct a junction with a single level in the central region is by creating an S-QD-S junction. This raises the question of the effect of interactions on the exceptional points. To study this, we model the system using the position and energy resolved Lindblad formalism\cite{kirsanskas_phenomenological_2018} (PERLind), which is valid for temperatures on the order of $\Gamma$\cite{nathan_universal_2020}. This allows us to capture the many-body nature of the problem, and we can compute the spectral function using the method in \cite{scarlatella_spectral_2019}(see Fig.~\ref{fig:U} and appendix). We see that the robust exceptional points are still present for small $U$.\\ While this does provide some indication that interaction effects do not destroy exceptional points, the PERLind method is limited in its applicability to temperatures on the order of $\Gamma$. At these temperatures, the signatures of the exceptional points are not visible in the differential conductance, as the thermal broadening washes out the features. However, the presence of the exceptional points in the spectral function does indicate that interactions do not automatically destroy the topological nature of the system. In addition, the fact that the coupling to the leads is strongly spin-polarized supresses the Kondo effect. Thus, while a more thorough treatment of the situation is required to make any definite statements, we do not expect that the situation changes at low temperatures.
	\\
	In conclusion, we have proposed a relatively simple experiment which allows for observation of signatures of exceptional points in multi-terminal SNS junctions. We have shown that this experimental setup can host different kinds of exceptional points, with some points marking topological phase boundaries and others marking symmetry-protected topological phase boundaries. Further, we have studied the behavior of these points in the case of a finite superconducting gap or Coulomb interaction, and shown that the topologically protected exceptional points are still present in these scenarios. This points towards the experimental feasibility of our setup, and towards the realization non-trivial non-Hermitian topology in multiterminal superconducting junctions.
	
	This research was funded in part by the European Research Council (Grant Agreement No. 856526) and by the DFG Collaborative Research Center (CRC) 183 Project No. 277101999.
	
%


\begin{thebibliography}{27}%
\makeatletter
\providecommand \@ifxundefined [1]{%
 \@ifx{#1\undefined}
}%
\providecommand \@ifnum [1]{%
 \ifnum #1\expandafter \@firstoftwo
 \else \expandafter \@secondoftwo
 \fi
}%
\providecommand \@ifx [1]{%
 \ifx #1\expandafter \@firstoftwo
 \else \expandafter \@secondoftwo
 \fi
}%
\providecommand \natexlab [1]{#1}%
\providecommand \enquote  [1]{``#1''}%
\providecommand \bibnamefont  [1]{#1}%
\providecommand \bibfnamefont [1]{#1}%
\providecommand \citenamefont [1]{#1}%
\providecommand \href@noop [0]{\@secondoftwo}%
\providecommand \href [0]{\begingroup \@sanitize@url \@href}%
\providecommand \@href[1]{\@@startlink{#1}\@@href}%
\providecommand \@@href[1]{\endgroup#1\@@endlink}%
\providecommand \@sanitize@url [0]{\catcode `\\12\catcode `\$12\catcode
  `\&12\catcode `\#12\catcode `\^12\catcode `\_12\catcode `\%12\relax}%
\providecommand \@@startlink[1]{}%
\providecommand \@@endlink[0]{}%
\providecommand \url  [0]{\begingroup\@sanitize@url \@url }%
\providecommand \@url [1]{\endgroup\@href {#1}{\urlprefix }}%
\providecommand \urlprefix  [0]{URL }%
\providecommand \Eprint [0]{\href }%
\providecommand \doibase [0]{https://doi.org/}%
\providecommand \selectlanguage [0]{\@gobble}%
\providecommand \bibinfo  [0]{\@secondoftwo}%
\providecommand \bibfield  [0]{\@secondoftwo}%
\providecommand \translation [1]{[#1]}%
\providecommand \BibitemOpen [0]{}%
\providecommand \bibitemStop [0]{}%
\providecommand \bibitemNoStop [0]{.\EOS\space}%
\providecommand \EOS [0]{\spacefactor3000\relax}%
\providecommand \BibitemShut  [1]{\csname bibitem#1\endcsname}%
\let\auto@bib@innerbib\@empty
\bibitem [{\citenamefont {{San-Jose}}\ \emph {et~al.}(2016)\citenamefont
  {{San-Jose}}, \citenamefont {Cayao}, \citenamefont {Prada},\ and\
  \citenamefont {Aguado}}]{san-jose_majorana_2016}%
  \BibitemOpen
  \bibfield  {author} {\bibinfo {author} {\bibfnamefont {P.}~\bibnamefont
  {{San-Jose}}}, \bibinfo {author} {\bibfnamefont {J.}~\bibnamefont {Cayao}},
  \bibinfo {author} {\bibfnamefont {E.}~\bibnamefont {Prada}},\ and\ \bibinfo
  {author} {\bibfnamefont {R.}~\bibnamefont {Aguado}},\ }\bibfield  {title}
  {\bibinfo {title} {Majorana bound states from exceptional points in
  non-topological superconductors},\ }\href {https://doi.org/10.1038/srep21427}
  {\bibfield  {journal} {\bibinfo  {journal} {Sci Rep}\ }\textbf {\bibinfo
  {volume} {6}},\ \bibinfo {pages} {21427} (\bibinfo {year}
  {2016})}\BibitemShut {NoStop}%
\bibitem [{\citenamefont {Ashida}\ \emph {et~al.}(2020)\citenamefont {Ashida},
  \citenamefont {Gong},\ and\ \citenamefont
  {Ueda}}]{ashida_non-hermitian_2020}%
  \BibitemOpen
  \bibfield  {author} {\bibinfo {author} {\bibfnamefont {Y.}~\bibnamefont
  {Ashida}}, \bibinfo {author} {\bibfnamefont {Z.}~\bibnamefont {Gong}},\ and\
  \bibinfo {author} {\bibfnamefont {M.}~\bibnamefont {Ueda}},\ }\bibfield
  {title} {\bibinfo {title} {Non-{{Hermitian}} physics},\ }\href
  {https://doi.org/10.1080/00018732.2021.1876991} {\bibfield  {journal}
  {\bibinfo  {journal} {Advances in Physics}\ }\textbf {\bibinfo {volume}
  {69}},\ \bibinfo {pages} {249} (\bibinfo {year} {2020})}\BibitemShut
  {NoStop}%
\bibitem [{\citenamefont {Wang}\ \emph {et~al.}(2023)\citenamefont {Wang},
  \citenamefont {Fu}, \citenamefont {Mao}, \citenamefont {Qie}, \citenamefont
  {Stone},\ and\ \citenamefont {Yang}}]{wang_non-hermitian_2023}%
  \BibitemOpen
  \bibfield  {author} {\bibinfo {author} {\bibfnamefont {C.}~\bibnamefont
  {Wang}}, \bibinfo {author} {\bibfnamefont {Z.}~\bibnamefont {Fu}}, \bibinfo
  {author} {\bibfnamefont {W.}~\bibnamefont {Mao}}, \bibinfo {author}
  {\bibfnamefont {J.}~\bibnamefont {Qie}}, \bibinfo {author} {\bibfnamefont
  {A.~D.}\ \bibnamefont {Stone}},\ and\ \bibinfo {author} {\bibfnamefont
  {L.}~\bibnamefont {Yang}},\ }\bibfield  {title} {\bibinfo {title}
  {Non-{{Hermitian}} optics and photonics: From classical to quantum},\ }\href
  {https://doi.org/10.1364/AOP.475477} {\bibfield  {journal} {\bibinfo
  {journal} {Adv. Opt. Photon., AOP}\ }\textbf {\bibinfo {volume} {15}},\
  \bibinfo {pages} {442} (\bibinfo {year} {2023})}\BibitemShut {NoStop}%
\bibitem [{\citenamefont {Zhou}\ \emph {et~al.}(2018)\citenamefont {Zhou},
  \citenamefont {Peng}, \citenamefont {Yoon}, \citenamefont {Hsu},
  \citenamefont {Nelson}, \citenamefont {Fu}, \citenamefont {Joannopoulos},
  \citenamefont {Solja{\v c}i{\'c}},\ and\ \citenamefont
  {Zhen}}]{zhou_observation_2018}%
  \BibitemOpen
  \bibfield  {author} {\bibinfo {author} {\bibfnamefont {H.}~\bibnamefont
  {Zhou}}, \bibinfo {author} {\bibfnamefont {C.}~\bibnamefont {Peng}}, \bibinfo
  {author} {\bibfnamefont {Y.}~\bibnamefont {Yoon}}, \bibinfo {author}
  {\bibfnamefont {C.~W.}\ \bibnamefont {Hsu}}, \bibinfo {author} {\bibfnamefont
  {K.~A.}\ \bibnamefont {Nelson}}, \bibinfo {author} {\bibfnamefont
  {L.}~\bibnamefont {Fu}}, \bibinfo {author} {\bibfnamefont {J.~D.}\
  \bibnamefont {Joannopoulos}}, \bibinfo {author} {\bibfnamefont
  {M.}~\bibnamefont {Solja{\v c}i{\'c}}},\ and\ \bibinfo {author}
  {\bibfnamefont {B.}~\bibnamefont {Zhen}},\ }\bibfield  {title} {\bibinfo
  {title} {Observation of bulk {{Fermi}} arc and polarization half charge from
  paired exceptional points},\ }\href {https://doi.org/10.1126/science.aap9859}
  {\bibfield  {journal} {\bibinfo  {journal} {Science}\ }\textbf {\bibinfo
  {volume} {359}},\ \bibinfo {pages} {1009} (\bibinfo {year}
  {2018})}\BibitemShut {NoStop}%
\bibitem [{\citenamefont {Ochkan}\ \emph {et~al.}(2024)\citenamefont {Ochkan},
  \citenamefont {Chaturvedi}, \citenamefont {K{\"o}nye}, \citenamefont
  {Veyrat}, \citenamefont {Giraud}, \citenamefont {Mailly}, \citenamefont
  {Cavanna}, \citenamefont {Gennser}, \citenamefont {Hankiewicz}, \citenamefont
  {B{\"u}chner}, \citenamefont {{van den Brink}}, \citenamefont {Dufouleur},\
  and\ \citenamefont {Fulga}}]{ochkan_non-hermitian_2024}%
  \BibitemOpen
  \bibfield  {author} {\bibinfo {author} {\bibfnamefont {K.}~\bibnamefont
  {Ochkan}}, \bibinfo {author} {\bibfnamefont {R.}~\bibnamefont {Chaturvedi}},
  \bibinfo {author} {\bibfnamefont {V.}~\bibnamefont {K{\"o}nye}}, \bibinfo
  {author} {\bibfnamefont {L.}~\bibnamefont {Veyrat}}, \bibinfo {author}
  {\bibfnamefont {R.}~\bibnamefont {Giraud}}, \bibinfo {author} {\bibfnamefont
  {D.}~\bibnamefont {Mailly}}, \bibinfo {author} {\bibfnamefont
  {A.}~\bibnamefont {Cavanna}}, \bibinfo {author} {\bibfnamefont
  {U.}~\bibnamefont {Gennser}}, \bibinfo {author} {\bibfnamefont {E.~M.}\
  \bibnamefont {Hankiewicz}}, \bibinfo {author} {\bibfnamefont
  {B.}~\bibnamefont {B{\"u}chner}}, \bibinfo {author} {\bibfnamefont
  {J.}~\bibnamefont {{van den Brink}}}, \bibinfo {author} {\bibfnamefont
  {J.}~\bibnamefont {Dufouleur}},\ and\ \bibinfo {author} {\bibfnamefont
  {I.~C.}\ \bibnamefont {Fulga}},\ }\bibfield  {title} {\bibinfo {title}
  {Non-{{Hermitian}} topology in a multi-terminal quantum {{Hall}} device},\
  }\href {https://doi.org/10.1038/s41567-023-02337-4} {\bibfield  {journal}
  {\bibinfo  {journal} {Nat. Phys.}\ }\textbf {\bibinfo {volume} {20}},\
  \bibinfo {pages} {395} (\bibinfo {year} {2024})}\BibitemShut {NoStop}%
\bibitem [{\citenamefont {Bergholtz}\ \emph {et~al.}(2021)\citenamefont
  {Bergholtz}, \citenamefont {Budich},\ and\ \citenamefont
  {Kunst}}]{bergholtz_exceptional_2021}%
  \BibitemOpen
  \bibfield  {author} {\bibinfo {author} {\bibfnamefont {E.~J.}\ \bibnamefont
  {Bergholtz}}, \bibinfo {author} {\bibfnamefont {J.~C.}\ \bibnamefont
  {Budich}},\ and\ \bibinfo {author} {\bibfnamefont {F.~K.}\ \bibnamefont
  {Kunst}},\ }\bibfield  {title} {\bibinfo {title} {Exceptional topology of
  non-{{Hermitian}} systems},\ }\href
  {https://doi.org/10.1103/RevModPhys.93.015005} {\bibfield  {journal}
  {\bibinfo  {journal} {Rev. Mod. Phys.}\ }\textbf {\bibinfo {volume} {93}},\
  \bibinfo {pages} {015005} (\bibinfo {year} {2021})}\BibitemShut {NoStop}%
\bibitem [{\citenamefont {Minganti}\ \emph {et~al.}(2020)\citenamefont
  {Minganti}, \citenamefont {Miranowicz}, \citenamefont {Chhajlany},
  \citenamefont {Arkhipov},\ and\ \citenamefont
  {Nori}}]{minganti_hybrid-liouvillian_2020}%
  \BibitemOpen
  \bibfield  {author} {\bibinfo {author} {\bibfnamefont {F.}~\bibnamefont
  {Minganti}}, \bibinfo {author} {\bibfnamefont {A.}~\bibnamefont
  {Miranowicz}}, \bibinfo {author} {\bibfnamefont {R.~W.}\ \bibnamefont
  {Chhajlany}}, \bibinfo {author} {\bibfnamefont {I.~I.}\ \bibnamefont
  {Arkhipov}},\ and\ \bibinfo {author} {\bibfnamefont {F.}~\bibnamefont
  {Nori}},\ }\bibfield  {title} {\bibinfo {title} {Hybrid-{{Liouvillian}}
  formalism connecting exceptional points of non-{{Hermitian Hamiltonians}} and
  {{Liouvillians}} via postselection of quantum trajectories},\ }\href
  {https://doi.org/10.1103/PhysRevA.101.062112} {\bibfield  {journal} {\bibinfo
   {journal} {Phys. Rev. A}\ }\textbf {\bibinfo {volume} {101}},\ \bibinfo
  {pages} {062112} (\bibinfo {year} {2020})}\BibitemShut {NoStop}%
\bibitem [{\citenamefont {Minganti}\ \emph {et~al.}(2019)\citenamefont
  {Minganti}, \citenamefont {Miranowicz}, \citenamefont {Chhajlany},\ and\
  \citenamefont {Nori}}]{minganti_quantum_2019}%
  \BibitemOpen
  \bibfield  {author} {\bibinfo {author} {\bibfnamefont {F.}~\bibnamefont
  {Minganti}}, \bibinfo {author} {\bibfnamefont {A.}~\bibnamefont
  {Miranowicz}}, \bibinfo {author} {\bibfnamefont {R.~W.}\ \bibnamefont
  {Chhajlany}},\ and\ \bibinfo {author} {\bibfnamefont {F.}~\bibnamefont
  {Nori}},\ }\bibfield  {title} {\bibinfo {title} {Quantum exceptional points
  of non-{{Hermitian Hamiltonians}} and {{Liouvillians}}: {{The}} effects of
  quantum jumps},\ }\href {https://doi.org/10.1103/PhysRevA.100.062131}
  {\bibfield  {journal} {\bibinfo  {journal} {Phys. Rev. A}\ }\textbf {\bibinfo
  {volume} {100}},\ \bibinfo {pages} {062131} (\bibinfo {year}
  {2019})}\BibitemShut {NoStop}%
\bibitem [{\citenamefont {M{\o}lmer}\ \emph {et~al.}(1993)\citenamefont
  {M{\o}lmer}, \citenamefont {Castin},\ and\ \citenamefont
  {Dalibard}}]{molmer_monte_1993}%
  \BibitemOpen
  \bibfield  {author} {\bibinfo {author} {\bibfnamefont {K.}~\bibnamefont
  {M{\o}lmer}}, \bibinfo {author} {\bibfnamefont {Y.}~\bibnamefont {Castin}},\
  and\ \bibinfo {author} {\bibfnamefont {J.}~\bibnamefont {Dalibard}},\
  }\bibfield  {title} {\bibinfo {title} {Monte {{Carlo}} wave-function method
  in quantum optics},\ }\href {https://doi.org/10.1364/JOSAB.10.000524}
  {\bibfield  {journal} {\bibinfo  {journal} {J. Opt. Soc. Am. B, JOSAB}\
  }\textbf {\bibinfo {volume} {10}},\ \bibinfo {pages} {524} (\bibinfo {year}
  {1993})}\BibitemShut {NoStop}%
\bibitem [{\citenamefont {Naghiloo}\ \emph {et~al.}(2019)\citenamefont
  {Naghiloo}, \citenamefont {Abbasi}, \citenamefont {Joglekar},\ and\
  \citenamefont {Murch}}]{naghiloo_quantum_2019}%
  \BibitemOpen
  \bibfield  {author} {\bibinfo {author} {\bibfnamefont {M.}~\bibnamefont
  {Naghiloo}}, \bibinfo {author} {\bibfnamefont {M.}~\bibnamefont {Abbasi}},
  \bibinfo {author} {\bibfnamefont {Y.~N.}\ \bibnamefont {Joglekar}},\ and\
  \bibinfo {author} {\bibfnamefont {K.~W.}\ \bibnamefont {Murch}},\ }\bibfield
  {title} {\bibinfo {title} {Quantum state tomography across the exceptional
  point in a single dissipative qubit},\ }\href
  {https://doi.org/10.1038/s41567-019-0652-z} {\bibfield  {journal} {\bibinfo
  {journal} {Nat. Phys.}\ }\textbf {\bibinfo {volume} {15}},\ \bibinfo {pages}
  {1232} (\bibinfo {year} {2019})}\BibitemShut {NoStop}%
\bibitem [{\citenamefont {Cayao}\ and\ \citenamefont
  {Sato}(2024{\natexlab{a}})}]{cayao_non-hermitian_2024}%
  \BibitemOpen
  \bibfield  {author} {\bibinfo {author} {\bibfnamefont {J.}~\bibnamefont
  {Cayao}}\ and\ \bibinfo {author} {\bibfnamefont {M.}~\bibnamefont {Sato}},\
  }\bibfield  {title} {\bibinfo {title} {Non-{{Hermitian}} multiterminal
  phase-biased {{Josephson}} junctions},\ }\href
  {https://doi.org/10.1103/PhysRevB.110.235426} {\bibfield  {journal} {\bibinfo
   {journal} {Phys. Rev. B}\ }\textbf {\bibinfo {volume} {110}},\ \bibinfo
  {pages} {235426} (\bibinfo {year} {2024}{\natexlab{a}})},\ \Eprint
  {https://arxiv.org/abs/2408.17260} {2408.17260} \BibitemShut {NoStop}%
\bibitem [{\citenamefont {Cayao}\ and\ \citenamefont
  {Sato}(2024{\natexlab{b}})}]{cayao_non-hermitian_2024-1}%
  \BibitemOpen
  \bibfield  {author} {\bibinfo {author} {\bibfnamefont {J.}~\bibnamefont
  {Cayao}}\ and\ \bibinfo {author} {\bibfnamefont {M.}~\bibnamefont {Sato}},\
  }\bibfield  {title} {\bibinfo {title} {Non-{{Hermitian}} phase-biased
  {{Josephson}} junctions},\ }\href
  {https://doi.org/10.1103/PhysRevB.110.L201403} {\bibfield  {journal}
  {\bibinfo  {journal} {Phys. Rev. B}\ }\textbf {\bibinfo {volume} {110}},\
  \bibinfo {pages} {L201403} (\bibinfo {year} {2024}{\natexlab{b}})},\ \Eprint
  {https://arxiv.org/abs/2307.15472} {2307.15472} \BibitemShut {NoStop}%
\bibitem [{\citenamefont {Li}\ \emph {et~al.}(2024)\citenamefont {Li},
  \citenamefont {Sun},\ and\ \citenamefont {Trauzettel}}]{li_anomalous_2024}%
  \BibitemOpen
  \bibfield  {author} {\bibinfo {author} {\bibfnamefont {C.-A.}\ \bibnamefont
  {Li}}, \bibinfo {author} {\bibfnamefont {H.-P.}\ \bibnamefont {Sun}},\ and\
  \bibinfo {author} {\bibfnamefont {B.}~\bibnamefont {Trauzettel}},\ }\bibfield
   {title} {\bibinfo {title} {Anomalous {{Andreev}} spectrum and transport in
  non-{{Hermitian Josephson}} junctions},\ }\href
  {https://doi.org/10.1103/PhysRevB.109.214514} {\bibfield  {journal} {\bibinfo
   {journal} {Phys. Rev. B}\ }\textbf {\bibinfo {volume} {109}},\ \bibinfo
  {pages} {214514} (\bibinfo {year} {2024})}\BibitemShut {NoStop}%
\bibitem [{\citenamefont {Ohnmacht}\ \emph {et~al.}(2024)\citenamefont
  {Ohnmacht}, \citenamefont {Wilhelm}, \citenamefont {Weisbrich},\ and\
  \citenamefont {Belzig}}]{ohnmacht_non-hermitian_2024}%
  \BibitemOpen
  \bibfield  {author} {\bibinfo {author} {\bibfnamefont {D.~C.}\ \bibnamefont
  {Ohnmacht}}, \bibinfo {author} {\bibfnamefont {V.}~\bibnamefont {Wilhelm}},
  \bibinfo {author} {\bibfnamefont {H.}~\bibnamefont {Weisbrich}},\ and\
  \bibinfo {author} {\bibfnamefont {W.}~\bibnamefont {Belzig}},\ }\href
  {https://doi.org/10.48550/arXiv.2408.01289} {\bibinfo {title} {Non-hermitian
  topology in multiterminal superconducting junctions}} (\bibinfo {year}
  {2024}),\ \Eprint {https://arxiv.org/abs/2408.01289} {2408.01289}
  \BibitemShut {NoStop}%
\bibitem [{\citenamefont {Shen}\ \emph {et~al.}(2024)\citenamefont {Shen},
  \citenamefont {Lu}, \citenamefont {Lado},\ and\ \citenamefont
  {Trif}}]{shen_non-hermitian_2024}%
  \BibitemOpen
  \bibfield  {author} {\bibinfo {author} {\bibfnamefont {P.-X.}\ \bibnamefont
  {Shen}}, \bibinfo {author} {\bibfnamefont {Z.}~\bibnamefont {Lu}}, \bibinfo
  {author} {\bibfnamefont {J.~L.}\ \bibnamefont {Lado}},\ and\ \bibinfo
  {author} {\bibfnamefont {M.}~\bibnamefont {Trif}},\ }\bibfield  {title}
  {\bibinfo {title} {Non-{{Hermitian Fermi-Dirac Distribution}} in {{Persistent
  Current Transport}}},\ }\href
  {https://doi.org/10.1103/PhysRevLett.133.086301} {\bibfield  {journal}
  {\bibinfo  {journal} {Phys. Rev. Lett.}\ }\textbf {\bibinfo {volume} {133}},\
  \bibinfo {pages} {086301} (\bibinfo {year} {2024})}\BibitemShut {NoStop}%
\bibitem [{\citenamefont {Capecelatro}\ \emph {et~al.}(2025)\citenamefont
  {Capecelatro}, \citenamefont {Marciani}, \citenamefont {Campagnano},\ and\
  \citenamefont {Lucignano}}]{capecelatro_andreev_2025}%
  \BibitemOpen
  \bibfield  {author} {\bibinfo {author} {\bibfnamefont {R.}~\bibnamefont
  {Capecelatro}}, \bibinfo {author} {\bibfnamefont {M.}~\bibnamefont
  {Marciani}}, \bibinfo {author} {\bibfnamefont {G.}~\bibnamefont
  {Campagnano}},\ and\ \bibinfo {author} {\bibfnamefont {P.}~\bibnamefont
  {Lucignano}},\ }\bibfield  {title} {\bibinfo {title} {Andreev non-{{Hermitian
  Hamiltonian}} for open {{Josephson}} junctions from {{Green}}'s functions},\
  }\href {https://doi.org/10.1103/PhysRevB.111.064517} {\bibfield  {journal}
  {\bibinfo  {journal} {Phys. Rev. B}\ }\textbf {\bibinfo {volume} {111}},\
  \bibinfo {pages} {064517} (\bibinfo {year} {2025})}\BibitemShut {NoStop}%
\bibitem [{\citenamefont {Pino}\ \emph {et~al.}(2025)\citenamefont {Pino},
  \citenamefont {Meir},\ and\ \citenamefont
  {Aguado}}]{pino_thermodynamics_2025}%
  \BibitemOpen
  \bibfield  {author} {\bibinfo {author} {\bibfnamefont {D.~M.}\ \bibnamefont
  {Pino}}, \bibinfo {author} {\bibfnamefont {Y.}~\bibnamefont {Meir}},\ and\
  \bibinfo {author} {\bibfnamefont {R.}~\bibnamefont {Aguado}},\ }\bibfield
  {title} {\bibinfo {title} {Thermodynamics of non-{{Hermitian Josephson}}
  junctions with exceptional points},\ }\href
  {https://doi.org/10.1103/PhysRevB.111.L140503} {\bibfield  {journal}
  {\bibinfo  {journal} {Phys. Rev. B}\ }\textbf {\bibinfo {volume} {111}},\
  \bibinfo {pages} {L140503} (\bibinfo {year} {2025})}\BibitemShut {NoStop}%
\bibitem [{\citenamefont {Javed}\ \emph {et~al.}(2023)\citenamefont {Javed},
  \citenamefont {Schwibbert},\ and\ \citenamefont
  {Riwar}}]{javed_fractional_2023}%
  \BibitemOpen
  \bibfield  {author} {\bibinfo {author} {\bibfnamefont {M.~A.}\ \bibnamefont
  {Javed}}, \bibinfo {author} {\bibfnamefont {J.}~\bibnamefont {Schwibbert}},\
  and\ \bibinfo {author} {\bibfnamefont {R.-P.}\ \bibnamefont {Riwar}},\
  }\bibfield  {title} {\bibinfo {title} {Fractional {{Josephson}} effect versus
  fractional charge in superconducting--normal metal hybrid circuits},\ }\href
  {https://doi.org/10.1103/PhysRevB.107.035408} {\bibfield  {journal} {\bibinfo
   {journal} {Phys. Rev. B}\ }\textbf {\bibinfo {volume} {107}},\ \bibinfo
  {pages} {035408} (\bibinfo {year} {2023})}\BibitemShut {NoStop}%
\bibitem [{\citenamefont {Pavlov}\ \emph {et~al.}(2025)\citenamefont {Pavlov},
  \citenamefont {Gefen},\ and\ \citenamefont
  {Shnirman}}]{pavlov_topological_2025}%
  \BibitemOpen
  \bibfield  {author} {\bibinfo {author} {\bibfnamefont {A.~I.}\ \bibnamefont
  {Pavlov}}, \bibinfo {author} {\bibfnamefont {Y.}~\bibnamefont {Gefen}},\ and\
  \bibinfo {author} {\bibfnamefont {A.}~\bibnamefont {Shnirman}},\ }\bibfield
  {title} {\bibinfo {title} {Topological transitions in quantum jump dynamics:
  {{Hidden}} exceptional points},\ }\href
  {https://doi.org/10.1103/PhysRevB.111.104301} {\bibfield  {journal} {\bibinfo
   {journal} {Phys. Rev. B}\ }\textbf {\bibinfo {volume} {111}},\ \bibinfo
  {pages} {104301} (\bibinfo {year} {2025})},\ \Eprint
  {https://arxiv.org/abs/2408.05270} {2408.05270} \BibitemShut {NoStop}%
\bibitem [{\citenamefont {Altland}\ and\ \citenamefont
  {Zirnbauer}(1997)}]{altland_nonstandard_1997}%
  \BibitemOpen
  \bibfield  {author} {\bibinfo {author} {\bibfnamefont {A.}~\bibnamefont
  {Altland}}\ and\ \bibinfo {author} {\bibfnamefont {M.~R.}\ \bibnamefont
  {Zirnbauer}},\ }\bibfield  {title} {\bibinfo {title} {Nonstandard symmetry
  classes in mesoscopic normal-superconducting hybrid structures},\ }\href
  {https://doi.org/10.1103/PhysRevB.55.1142} {\bibfield  {journal} {\bibinfo
  {journal} {Phys. Rev. B}\ }\textbf {\bibinfo {volume} {55}},\ \bibinfo
  {pages} {1142} (\bibinfo {year} {1997})}\BibitemShut {NoStop}%
\bibitem [{\citenamefont {Kawabata}\ \emph {et~al.}(2019)\citenamefont
  {Kawabata}, \citenamefont {Shiozaki}, \citenamefont {Ueda},\ and\
  \citenamefont {Sato}}]{kawabata_symmetry_2019}%
  \BibitemOpen
  \bibfield  {author} {\bibinfo {author} {\bibfnamefont {K.}~\bibnamefont
  {Kawabata}}, \bibinfo {author} {\bibfnamefont {K.}~\bibnamefont {Shiozaki}},
  \bibinfo {author} {\bibfnamefont {M.}~\bibnamefont {Ueda}},\ and\ \bibinfo
  {author} {\bibfnamefont {M.}~\bibnamefont {Sato}},\ }\bibfield  {title}
  {\bibinfo {title} {Symmetry and {{Topology}} in {{Non-Hermitian Physics}}},\
  }\href {https://doi.org/10.1103/PhysRevX.9.041015} {\bibfield  {journal}
  {\bibinfo  {journal} {Phys. Rev. X}\ }\textbf {\bibinfo {volume} {9}},\
  \bibinfo {pages} {041015} (\bibinfo {year} {2019})}\BibitemShut {NoStop}%
\bibitem [{\citenamefont {Beenakker}(2024)}]{beenakker_josephson_2024}%
  \BibitemOpen
  \bibfield  {author} {\bibinfo {author} {\bibfnamefont {C.~W.~J.}\
  \bibnamefont {Beenakker}},\ }\bibfield  {title} {\bibinfo {title} {Josephson
  effect in a junction coupled to an electron reservoir},\ }\href
  {https://doi.org/10.1063/5.0215522} {\bibfield  {journal} {\bibinfo
  {journal} {Applied Physics Letters}\ }\textbf {\bibinfo {volume} {125}},\
  \bibinfo {pages} {122601} (\bibinfo {year} {2024})},\ \Eprint
  {https://arxiv.org/abs/2404.13976} {2404.13976} \BibitemShut {NoStop}%
\bibitem [{\citenamefont {Budich}\ \emph {et~al.}(2019)\citenamefont {Budich},
  \citenamefont {Carlstr{\"o}m}, \citenamefont {Kunst},\ and\ \citenamefont
  {Bergholtz}}]{budich_symmetry-protected_2019}%
  \BibitemOpen
  \bibfield  {author} {\bibinfo {author} {\bibfnamefont {J.~C.}\ \bibnamefont
  {Budich}}, \bibinfo {author} {\bibfnamefont {J.}~\bibnamefont
  {Carlstr{\"o}m}}, \bibinfo {author} {\bibfnamefont {F.~K.}\ \bibnamefont
  {Kunst}},\ and\ \bibinfo {author} {\bibfnamefont {E.~J.}\ \bibnamefont
  {Bergholtz}},\ }\bibfield  {title} {\bibinfo {title} {Symmetry-protected
  nodal phases in non-{{Hermitian}} systems},\ }\href
  {https://doi.org/10.1103/PhysRevB.99.041406} {\bibfield  {journal} {\bibinfo
  {journal} {Phys. Rev. B}\ }\textbf {\bibinfo {volume} {99}},\ \bibinfo
  {pages} {041406} (\bibinfo {year} {2019})}\BibitemShut {NoStop}%
\bibitem [{\citenamefont {Kir{\v s}anskas}\ \emph {et~al.}(2018)\citenamefont
  {Kir{\v s}anskas}, \citenamefont {Francki{\'e}},\ and\ \citenamefont
  {Wacker}}]{kirsanskas_phenomenological_2018}%
  \BibitemOpen
  \bibfield  {author} {\bibinfo {author} {\bibfnamefont {G.}~\bibnamefont
  {Kir{\v s}anskas}}, \bibinfo {author} {\bibfnamefont {M.}~\bibnamefont
  {Francki{\'e}}},\ and\ \bibinfo {author} {\bibfnamefont {A.}~\bibnamefont
  {Wacker}},\ }\bibfield  {title} {\bibinfo {title} {Phenomenological position
  and energy resolving {{Lindblad}} approach to quantum kinetics},\ }\href
  {https://doi.org/10.1103/PhysRevB.97.035432} {\bibfield  {journal} {\bibinfo
  {journal} {Phys. Rev. B}\ }\textbf {\bibinfo {volume} {97}},\ \bibinfo
  {pages} {035432} (\bibinfo {year} {2018})}\BibitemShut {NoStop}%
\bibitem [{\citenamefont {Nathan}\ and\ \citenamefont
  {Rudner}(2020)}]{nathan_universal_2020}%
  \BibitemOpen
  \bibfield  {author} {\bibinfo {author} {\bibfnamefont {F.}~\bibnamefont
  {Nathan}}\ and\ \bibinfo {author} {\bibfnamefont {M.~S.}\ \bibnamefont
  {Rudner}},\ }\bibfield  {title} {\bibinfo {title} {Universal {{Lindblad}}
  equation for open quantum systems},\ }\href
  {https://doi.org/10.1103/PhysRevB.102.115109} {\bibfield  {journal} {\bibinfo
   {journal} {Phys. Rev. B}\ }\textbf {\bibinfo {volume} {102}},\ \bibinfo
  {pages} {115109} (\bibinfo {year} {2020})},\ \Eprint
  {https://arxiv.org/abs/2004.01469} {2004.01469} \BibitemShut {NoStop}%
\bibitem [{\citenamefont {Scarlatella}\ \emph {et~al.}(2019)\citenamefont
  {Scarlatella}, \citenamefont {Clerk},\ and\ \citenamefont
  {Schiro}}]{scarlatella_spectral_2019}%
  \BibitemOpen
  \bibfield  {author} {\bibinfo {author} {\bibfnamefont {O.}~\bibnamefont
  {Scarlatella}}, \bibinfo {author} {\bibfnamefont {A.~A.}\ \bibnamefont
  {Clerk}},\ and\ \bibinfo {author} {\bibfnamefont {M.}~\bibnamefont
  {Schiro}},\ }\bibfield  {title} {\bibinfo {title} {Spectral functions and
  negative density of states of a driven-dissipative nonlinear quantum
  resonator},\ }\href {https://doi.org/10.1088/1367-2630/ab0ce9} {\bibfield
  {journal} {\bibinfo  {journal} {New J. Phys.}\ }\textbf {\bibinfo {volume}
  {21}},\ \bibinfo {pages} {043040} (\bibinfo {year} {2019})}\BibitemShut
  {NoStop}%
\bibitem [{\citenamefont {Stefanucci}\ and\ \citenamefont {{van
  Leeuwen}}(2013)}]{stefanucci_nonequilibrium_2013}%
  \BibitemOpen
  \bibfield  {author} {\bibinfo {author} {\bibfnamefont {G.}~\bibnamefont
  {Stefanucci}}\ and\ \bibinfo {author} {\bibfnamefont {R.}~\bibnamefont {{van
  Leeuwen}}},\ }\href {https://doi.org/10.1017/CBO9781139023979} {\emph
  {\bibinfo {title} {Nonequilibrium {{Many-Body Theory}} of {{Quantum
  Systems}}: {{A Modern Introduction}}}}}\ (\bibinfo  {publisher} {Cambridge
  University Press},\ \bibinfo {address} {Cambridge},\ \bibinfo {year}
  {2013})\BibitemShut {NoStop}%
\end{thebibliography}
	
	\appendix
	\onecolumngrid
	\section{Transport formalism}
	We describe the system as consisting of a central region, $N$ leads, and a tunneling between them:
	\begin{equation}
		H=H_C+H_L+H_T.
	\end{equation}
	Here, 
	\begin{equation}
		H_T=\sum_{n,i,p}(d_{n}^{\dagger}T_{n,ip}c_{ip}+h.c),
	\end{equation}
	where $d_{n}=(d_{n\uparrow},d_{n\downarrow},d_{n\downarrow}^\dagger,-d_{n\uparrow}^\dagger)$ is the Nambu spinor of the central region, $c_{ip}=(c_{ip\uparrow},c_{ip\downarrow},c_{ip\downarrow}^\dagger,-c_{ip\uparrow}^\dagger)$ is the Nambu spinor of lead $i$, and $T_{n,ip}$ is a tunneling matrix. The current into lead $i$ can be written as
	\begin{equation}
		I_i=ie\sum_{n,p}(d_{n}^{\dagger}T_{n,ip}\tau_3 c_{ip}-c_{ip}^\dagger T_{n,ip}^\dagger\tau_3 d_n),
	\end{equation}
	where $\tau_n$ are the Pauli matrices in Nambu space. In the language of non-equilibrium Green functions, we can write the expectation value of the current as
	\begin{equation}
		\langle I_i\rangle=-2e\Re\sum_{n,p}\Tr[G^<(n,p,i,t,t)T^*_{n,ip}],
	\end{equation}
	where $G(n,p,i,t,t')=-i\langle \mathcal{T}d_n^\dagger(t)c_{ip}(t)\rangle$ is the mixed Nambu Green function. We can now use the Langreth rules\cite{stefanucci_nonequilibrium_2013} to write this in terms of the center and lead Green functions,
	\begin{equation}
		\langle I_i\rangle=2e\int\frac{d\omega}{2\pi}\sum_{n'}[G^R(n,n',\omega)\Sigma_0^<(n',n,i,\omega)\tau_3+G^<(n,n',\omega)\Sigma_0^A(n',n,i,\omega)\tau_3],
		\label{Current}
	\end{equation}
	where $G(n,n',\omega)$ is the center Green function, and we define the self-energies as
	\begin{equation}
		\Sigma_0(n',n,i,\omega)=\sum_pT_{n',ip}G_0(p,i,\omega)T_{n,ip},
	\end{equation}
	where $G_0(p,i,\omega)$ is the Green function of lead $i$. We can now compute the self-energies of the different leads. For the normal leads, we get
	\begin{equation*}
		\Sigma^R(n',n,i,\omega)=\sum_pT_{n',ip}\frac{1}{\omega-\tau_3\xi_p+i0^+}T_{n,ip}
	\end{equation*}
	and for the superconducting leads, we get
	\begin{equation*}
		\Sigma^R(n',n,i,\omega)=\sum_pT_{n',ip}\frac{1}{(\omega+i0^+)^2-E_p^2}\begin{pmatrix}
			\omega+\xi_p & 0 & -\Delta_i&0 \\ 0 & \omega+\xi_p & 0 & -\Delta_i \\ -\Delta_i^* & 0 & \omega-\xi_p & 0 \\ 0 & -\Delta_i^* & 0 & \omega-\xi_p
		\end{pmatrix}T_{n,ip},
	\end{equation*}
	where $E_p=\sqrt{\xi_p^2+\abs{\Delta}^2}$. In the wide-band limit, these sums can be performed, yielding
	\begin{equation}
		\Sigma_R(n',n,\mathrm{normal},\omega)=-i\frac{\mathbf{\Gamma}_{n,n'}}{2}
	\end{equation}
	where $\mathbf{\Gamma}_{n,n'}$ is the dissipative coupling, which is a hermitian matrix, and
	\begin{equation}
		\Sigma_R(n',n,SC,\omega)=\frac{i\gamma}{(\omega+i0^+)^2-\abs{\Delta}^2}\begin{pmatrix}
			\omega & 0 & \Delta & 0\\0 & \omega & 0 & \Delta\\ \Delta & 0 & \omega & 0 \\ 0 & \Delta & 0 & \omega
		\end{pmatrix}
	\end{equation}
	The advanced self-energies are calculated through $\Sigma^A=(\Sigma^R)^*$, or alternatively through $\Sigma(\omega-i0^+)\rightarrow\Sigma(\omega+i0^+)$. The lesser self-energies are calculated through $\Sigma^<(\omega)=(\Sigma^R(\omega)-\Sigma^A(\omega))n_f(\omega-\tau_3eV)$. Finally, the lesser Green function can be computed as $G^<=G^R\Sigma^<G^A$. The current is then calculated using equation (\ref{Current}). The spectral function is also easily computed, since $A(\omega)=2\Im(G^R(\omega))$.
	\section{PERLind}
	\renewcommand{\theequation}{B\arabic{equation}}
	\setcounter{equation}{0}
	To compute the spectral function in the interacting case, we use the position and energy resolved Lindblad method (PERLind)\cite{kirsanskas_phenomenological_2018}. Under the assumption that the tunneling process is Markovian, equivalent to the assumption that the temperature is at least of the same order of magnitude as the tunneling coefficient, the PERLind method is a consistent method for finding the time evolution of a system coupled to a number of leads. The method allows one to define a Lindblad operator of the form
	\begin{equation}
		\dot{\rho}=\mathcal{L}\rho=-i[H,\rho]+\sum_iL_i^\dagger\rho L_i-\frac{1}{2}\{L_i^\dagger L_i,\rho\},
	\end{equation}
	where $H$ is the many-body Hamiltonian of the central region, and $L_i$ are jump operators modeling tunneling between the central region and the normal leads. The many-body Hamiltonian includes the proximitization of the superconducting leads, as well as the Coulomb repulsion. It is defined in the usual way. The key idea of the PERLind method is that it allows one to define the jump operators in a consistent way. To define these jump operators, we start by identifying the types of jumps $\hat{L}_i$ which are present. In our case, this means tunneling of spin-up and spin-down electrons into and out of the central region from the normal lead and the probe lead. An example jump operator could thus be $\Gamma_\uparrow d^\dagger_\uparrow$ for a spin-up electron tunneling into the central region from the normal lead. We then go to the diagonal basis of $H$ and weigh the jump operators by a Fermi factor $L_i=\hat{L}_i\sqrt{f(\Delta E-eV_j)}$ where $\Delta E_{mn}=E_m-E_n$ is a matrix encoding the energy difference between the state before and after the jump (Since we are in a basis diagonal in $H$, these energies are all well-defined) and $V_i$ is the voltage of the $j$'th lead.\\
	Once the jump operators have been defined, the Lindbladian can be written and diagonalized numerically, and one can find the eigenvalues $\lambda_i$ and corresponding left and right eigenoperators $l_i$ and $r_i$. In particular, one can find the steady-state density operator $\rho_s$, defined as the right eigenoperator whose corresponding left eigenoperator is the identity. The probe current can now be calculated as
	\begin{equation}
		I=\sum_{i}N\cdot\left(L_i\rho_s L_i^\dagger-\frac{1}{2}\{L_i^\dagger L_i,\rho_s\}\right),
	\end{equation}
	where $i$ runs over all of the probe jump operators, $N$ is the particle number operator, and $\rho_s$ is the steady-state density matrix of the system.\\
	To compute the spectral function, the method in \cite{scarlatella_spectral_2019} is used. This method prescribes that the retarded Green function can be written as
	\begin{equation}
		G^R(\omega)=\sum_\alpha\frac{w_\alpha}{\omega-i\lambda_\alpha}
	\end{equation}
	where $\alpha$ runs over the number of eigenvalues of $\mathcal{L}$ and $w_\alpha=\sum_i\Tr{\hat{L}_i r_\alpha}\Tr{l_\alpha^\dagger[\hat{L}_i^\dagger,\rho_s]}$. The spectral function is then calculated as the imaginary part of $G^R$. This allows us to compute both the current and the spectral function. At the relevant temperatures $T\sim\Gamma$, however, the thermal broadening of the current as a function of $V$ makes it impossible to see any features of the exceptional point.
	\section{Robust exceptional point}\label{robust-EP}
	\renewcommand{\theequation}{C\arabic{equation}}
	\setcounter{equation}{0}
	To study the robust exceptional points, we start with the non-hermitian effective Hamiltonian of the system given by (\ref{hamiltonian}). We perform a rotation such that the $z$-direction is the direction of $\vec{B}$, and the dissipation is off-diagonal, yielding
	\begin{equation}
		\tilde{H}=\begin{pmatrix}
			\tilde{\varepsilon}_\uparrow & i\Gamma(\theta) & \gamma\cos{\frac{\phi}{2}} & 0\\
			i\Gamma(\theta) & \tilde{\varepsilon}_\downarrow & 0 & \gamma\cos{\frac{\phi}{2}}\\
			\gamma\cos{\frac{\phi}{2}} & 0 & -\tilde{\varepsilon}_\downarrow^* & i\Gamma(\theta)\\
			0 & \gamma\cos{\frac{\phi}{2}} & i\Gamma(\theta) & -\tilde{\varepsilon}_\uparrow^*
		\end{pmatrix}
	\end{equation}
	where $i\Gamma(\theta)\propto\cos{\theta}$. For $\Gamma(\theta)=0$, this matrix can be written in a block-diagonal form and diagonalized by a Bogoliubov transformation. We can perform the transformation even if $\Gamma(\theta)\neq0$, yielding
	\begin{equation}
		\tilde{H}=\begin{pNiceArray}{cc|cc}
			\varepsilon_1(\phi) & 0 & \Block{2-2}<\Large>{i\mathbf{K}} \\
			0 & -\varepsilon_2^*(\phi) \\
			\hline
			\Block{2-2}<\Large>{i\mathbf{K}^\dagger} && \varepsilon_2(\phi) & 0 \\
			&& 0 & -\varepsilon_1^*(\phi)
		\end{pNiceArray}
	\end{equation}
	where $\mathbf{K}$ is a matrix proportional to $\cos{\theta}$. The subspace described by (\ref{robust}) is then simply the subspace given by $\varepsilon_1$ and $-\varepsilon_1^*$, or equivalently for $\varepsilon_2$.
\end{document}